\documentclass[final]{svjour3}

\usepackage{graphicx,subfigure}
\usepackage{physics}
\usepackage{rotating}
\usepackage{amssymb}
\usepackage{mathptmx}
\usepackage[numbers,compress]{natbib}
\usepackage{siunitx}
\usepackage{verbatim}
\usepackage{multirow}
\usepackage{threeparttable}
\usepackage{upgreek}

\usepackage{xcolor}
\usepackage{ulem}

\makeatletter
\journalname{Journal of Low Temperature Physics}

\graphicspath{ {figs/} }

\begin{document}

\newcommand{\hdblarrow}{H\makebox[0.9ex][l]{$\downdownarrows$}-}
\newcommand{\CMO}{CaMoO$_4$}
\newcommand{\CWO}{CaWO$_4$}
\newcommand{\enCMO}{$^\mathrm{48depl.}$Ca$^{100}$MoO$_4$}
\newcommand{\zerodbd}{$0\nu\beta\beta$}
\newcommand{\twodbd}{$2\nu\beta\beta$}
\newcommand{\LMO}{Li$_2$MoO$_4$}
\newcommand{\enLMO}{Li$_2$$^{100}$MoO$_4$}

\newcommand{\mbb}{$ m_{\beta\beta}$}
\newcommand{\Tzerov}{$T_{1/2}^{0\nu}$}
\newcommand{\gA}{$g_\mathrm{A}$}
\newcommand{\gAeff}{$g_\mathrm{A}^\mathrm{eff}$}

\newcommand{\alp}{$\alpha$}
\newcommand{\bg}{$\beta$/$\gamma$}

\newcommand{\Mo}[1]{$^{#1}$Mo}
\newcommand{\Fe}[1]{$^{#1}$Fe}

\title{Status and performance of the AMoRE-I experiment on neutrinoless double beta decay}

\author{H.B.~Kim$^{1,2}$ \and D.H.~Ha$^{1}$ \and E.J.~Jeon$^{1,3}$ 
\and
J.A.~Jeon$^{1}$ \and H.S.~Jo$^{1}$ \and C.S.~Kang$^{1}$ \and W.G.~Kang$^{1}$ \and H.S.~Kim$^{4}$ \and S.C.~Kim$^{1}$ \and S.G.~Kim$^{1}$ \and S.K.~Kim$^{2}$  \and S.R.~Kim$^{1}$ \and W.T.~Kim$^{1,3}$ \and Y.D.~Kim$^{1,3}$ \and Y.H.~Kim$^{1,3,5}$ \and D.H.~Kwon$^{1,3}$ \and E.S.~Lee$^{1}$ \and H.J.~Lee$^{1}$ \and H.S.~Lee$^{1,3}$ \and J.S.~Lee$^{1}$ \and M.H.~Lee$^{1,3}$ \and S.W.~Lee$^{1}$ \and Y.C.~Lee$^{1,2}$ 
\and D.S.~Leonard$^{1}$
\and H.S.~Lim$^{1}$ \and B.~Mailyan$^{1}$ \and P.B.~Nyanda$^{1,3}$ \and Y.M.~Oh$^{1}$ \and M.B.~Sari$^{6}$ \and J.W.~Seo$^{1,3}$ \and K.M.~Seo$^{1}$ \and S.H.~Seo$^{1}$ \and J.H.~So$^{7}$ \and K.R.~Woo$^{1,3}$ \and Y.S.~Yoon$^{5}$}

\institute{
$^{1}$ Center for Underground Physics, Institute for Basic Science (IBS), Daejeon 34126, Korea,
\\$^{2}$ Department of Physics and Astronomy, Seoul National University, Seoul 08826, Korea,
\\$^{3}$ IBS School, University of Science and Technology (UST), Daejeon 34113, Korea,
\\$^{4}$ Department of Physics and Astronomy, Sejong University, Seoul 05006, Korea,
\\$^{5}$ Korea Research Institute of Standards and Science (KRISS), Daejeon 34114, Korea,
\\$^{6}$ Faculty of Mathematics and Natural Science, Bandung Institute of Technology, Bandung 40132, Indonesia,
\\$^{7}$ Department of Physics, Kyungpook National University, Daegu 41566, Korea,
\\  \email{hanbum7@snu.ac.kr, skkim@snu.ac.kr, yhk@ibs.re.kr}}

\maketitle

\begin{abstract}
AMoRE is an international project to search for the neutrinoless double beta decay of \Mo{100} using a detection technology consisting of magnetic microcalorimeters (MMCs) and molybdenum-based scintillating crystals.
Data collection has begun for the current AMORE-I phase of the project, an upgrade from the previous pilot phase. 
AMoRE-I employs thirteen \enCMO{} crystals and five \enLMO{} crystals for a total crystal mass of 6.2\,kg. Each detector module contains a scintillating crystal with two MMC channels for heat and light detection. 
We report the present status of the experiment and the performance of the detector modules. 

\end{abstract}

\section{Introduction}
AMoRE is an international project to search for neutrinoless double beta (\zerodbd) decay of \Mo{100} using low-temperature heat and light detection by Mo-based scintillating crystals~\cite{amore_tdr}. We employ enriched molybdate crystals as both source and detection materials for $0\nu\beta\beta$~decay of \Mo{100}. A high natural abundance of 9.7\% and a Q-value of 3.034\,MeV that is greater than most environmental $\gamma$ lines makes \Mo{100} a competitive isotope in the search for \zerodbd{} decay.
The readout scheme of the detector channels is based on magnetic microcalorimeters (MMCs), which can provide fast response and high sensitivity in the operating temperature regime of tens of mK. MMCs offer not only high resolution in time and energy~\cite{CPD,SALCJ}, which is essential for discriminating events by time characteristics, but also high linearity in energy and a wide dynamic range enabling reliable energy calibration and multitemperature testing.

During 2016--2018, the AMoRE-pilot experiment was carried out utilizing six crystals of \enCMO{} with a total mass of 1.9 kg~\cite{hsjo2018,pilot_result}.

The pilot experiment demonstrated a \zerodbd{} sensitivity of $T_{1/2}=3.43\times10^{23}$\,yr at the 90\% confidence level. Through the pilot experiment, we demonstrated that the MMC-based low-temperature detector technology and enriched molybdate crystals were suitable for a large-scale long-term experiment with competitive detection sensitivity.

AMoRE-I is the current phase of the AMoRE project that has been running stably with the same dilution refrigerator used in the pilot phase at the 700-meter-deep Yangyang Underground Laboratory (Y2L).
In this report we present the status and performance of AMoRE-I.
The AMoRE-I detector is a 6.2\,kg array of thirteen \enCMO{} and five \enLMO{} crystals including those used in the pilot phase. As an independent physics experiment, AMoRE-I aims at the background level in the order of 0.01 counts/keV/kg/year at $Q_{\beta\beta}$which equates to a half-life limit of $\sim10^{24}$ years. AMoRE-I sensitivity will be comparable to the recent half-lift limit of 1.8$\times$10$^{24}$\,years found in the CUPID-Mo project using \enLMO{} crystals and NTD Ge thermistors~\cite{cupidmo2022}.

An additional objective of AMoRE-I is to perform further research and development on the detector and to provide background for the next phase of the AMoRE experiment, AMoRE-II, in which measurements will be performed using a total mass of approximately 100\,kg of \Mo{100}.

\begin{figure}[t]
\begin{center}
\includegraphics[width=0.45\linewidth, keepaspectratio]{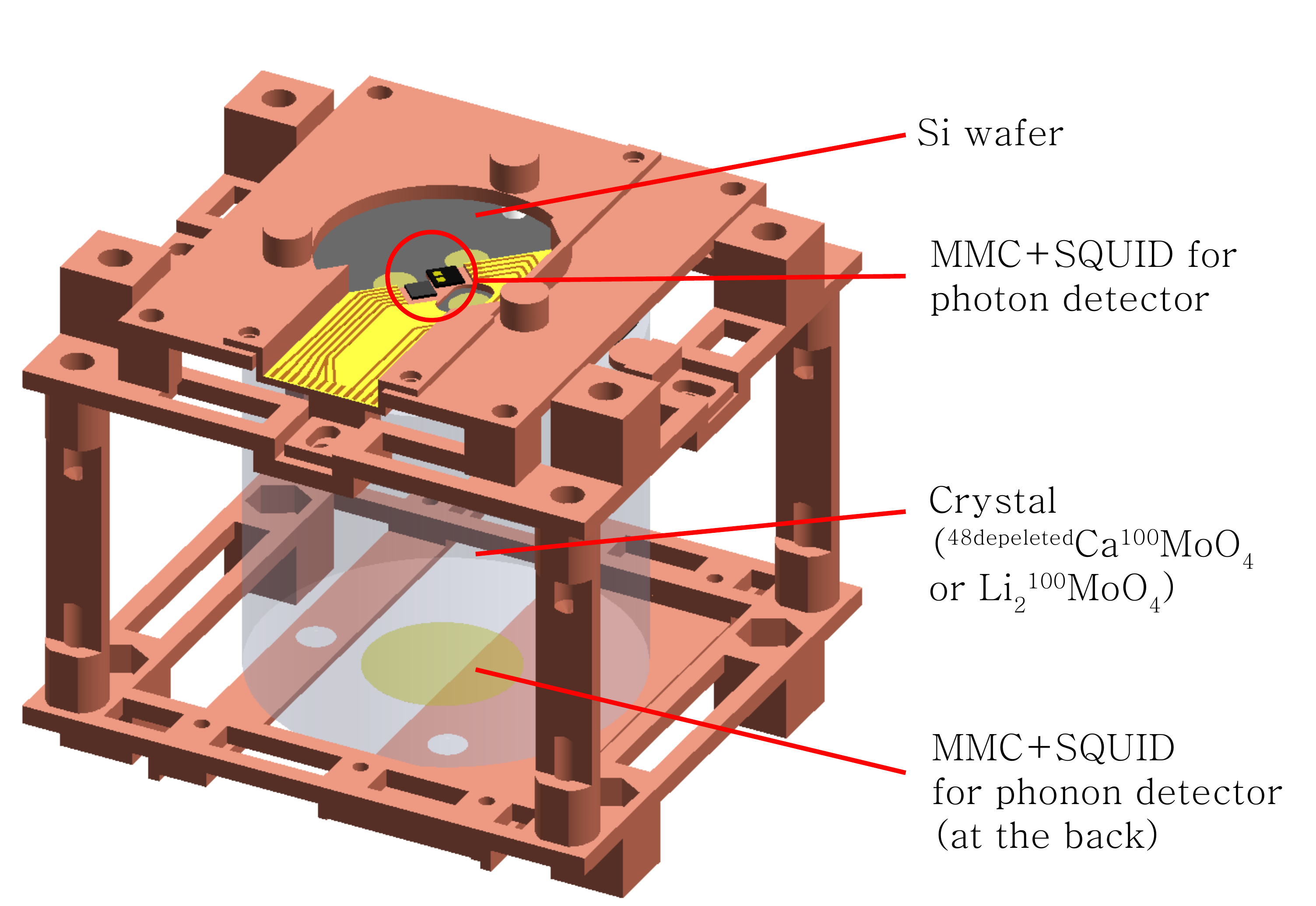}
\includegraphics[width=0.22\linewidth, keepaspectratio]{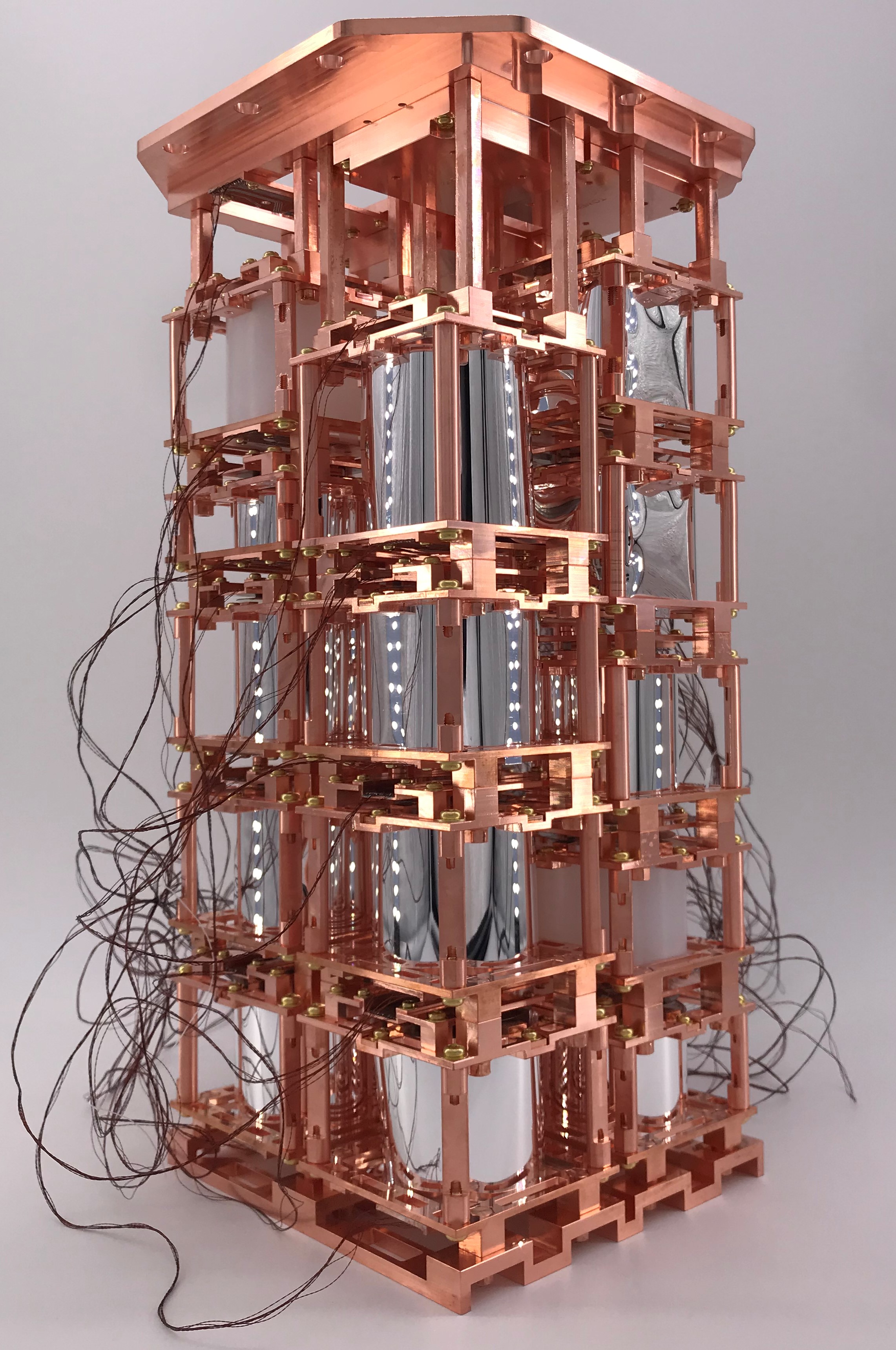}
\label{fig:modules}
\caption{Detector module used in AMoRE-I ({\it left}). The detector is composed of an \Mo{100}-enriched crystal, a photon detector, and a phonon detector. The detector tower of AMoRE-I consisting 18 detector modules in four columns ({\it right}).} 
\end{center}
\end{figure}

\section{Experimental setup}

A single AMoRE-I detector module is comprised of a molybdate scintillating crystal whose size varies $\Phi \geq 4$ cm / H $\lesssim$ 5 cm, and a pair of phonon and photon detectors each of which includes an MMC sensor used for the temperature readout in thermal calorimetric detection. 
The simultaneous detection of phonon and photon signals enables us to discriminate $\alpha$ backgrounds from $\beta/\gamma$ events~\cite{gbkim2017,iwkim2017}. 

Energy absorption by a target absorber of a scintillating crystal causes a temperature increase in the crystal that is measured with the MMC sensor used for the phonon channel. A light detector composed of a Si or Ge wafer and another MMC sensor is employed to measure the scintillation photons emitted from the target crystal. 

Fig.~\ref{fig:modules} shows the design used for all the AMoRE-I detector modules. 
A target crystal is held by PTFE clamps (not shown in Fig.~\ref{fig:modules}) in a frame structure made of low-background copper. This copper frame serves as a cooling path for the crystals and sensors, as well as a thermal bath for the detector module.
The MMC sensor for the phonon channel is located at the bottom of the module design. 
The crystal has a vapor-deposited gold film on its bottom surface. 
A thermal connection between the phonon-collector film and the MMC sensor is formed using twenty-four annealed gold wires.
An additional gold wire is attached to the MMC sensor and the copper holder to create a thermal connection to the heat bath.
In this structure, the heat capacity of the MMC sensor dominates that of the detector at low-temperature scale~\cite{sgkim2021}.

Considering the hygroscopic characteristics of \LMO{}, we developed a strict protocol for the cleaning procedure before vapor deposition to be performed after polishing the crystal surface. 
We ground one surface of the \LMO{} crystal using 1500-grit SiC sandpaper to remove any material damaged by moisture adsorption and to ensure that the phonon collector film adhered to the \LMO{} surface. A 14-mm diameter circular film of Ti/Au with a thickness of 5/300\,nm was evaporated onto the ground surface in 50-mm diameter of the \LMO{} crystal. One of five \LMO{} crystals, coated with a metal film of Ti/Ag/Au with thicknesses of 5/300/10\,nm, was used to perform various material studies on phonon collector films.

\color{black}
A light detector was placed on top of the crystal as shown in Fig.~\ref{fig:modules}. 
The crystal was surrounded by light-reflection foils with high reflectivity in order to increase the collection efficiency of the scintillation photons. Two-inch Si wafers with a 90-nm-thick SiO$_2$ anti-reflection coating constituted the absorber of the light detector. The wafer was equipped with three 3-mm-diameter phonon collector films that were used with bonding wires to form a thermal connection to the MMC photon sensor~\cite{hjlee2015,sgkim}.
 
All the phonon channels and 10 photon channels were equipped with MMC sensors from two AgEr batches with concentrations of 299\,ppm and 414\,ppm~\cite{sgkim2021}.
The other photon channels employed the MMC used during the pilot phase.
Moreover, on the flat surface of each crystal where the metal film is deposited, a Si heater is attached for stabilization~\cite{dhkwon2020}.

\begin{figure}[t]
\begin{center}
\includegraphics[width=0.5\linewidth, keepaspectratio]{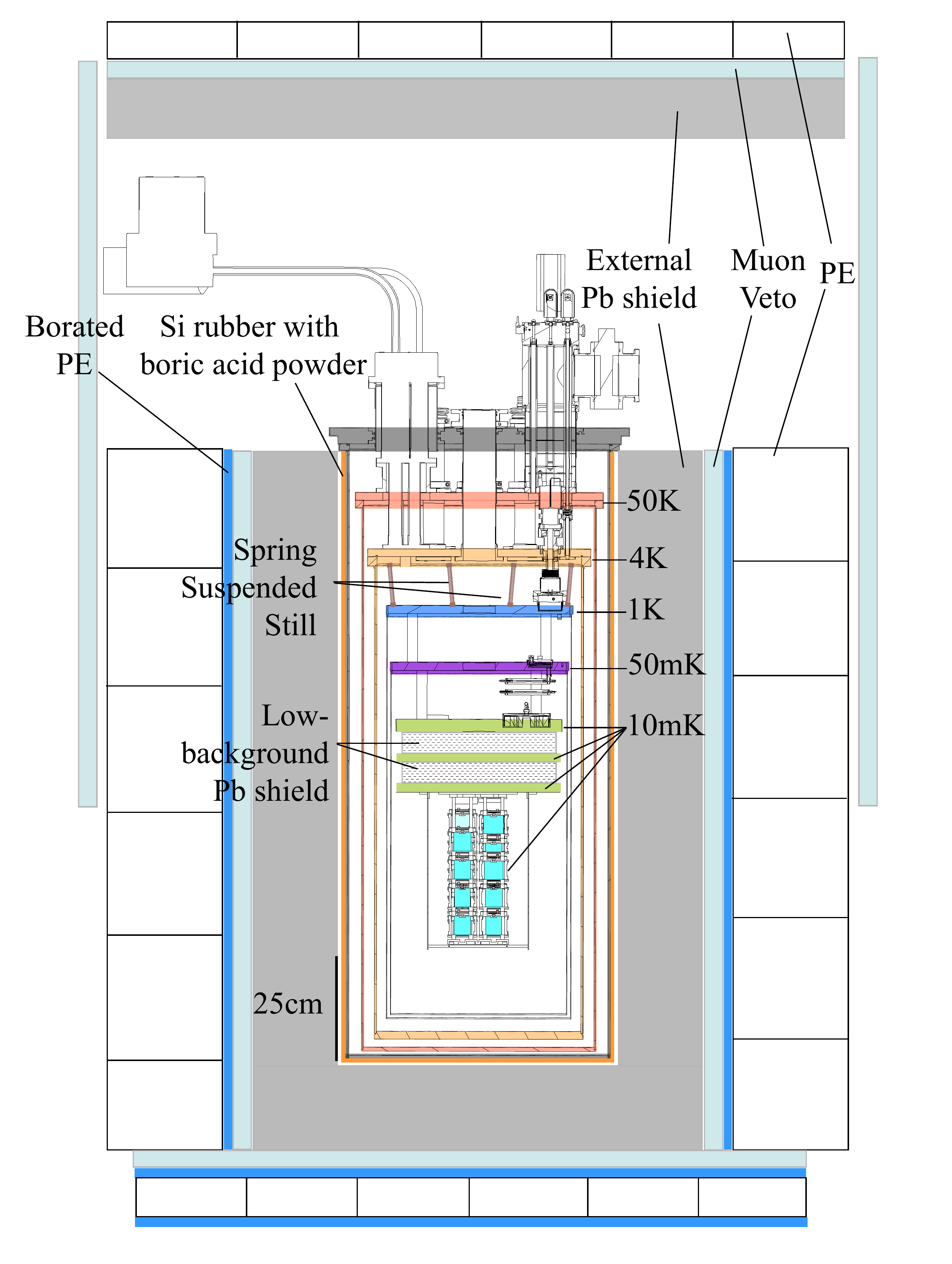}
\caption{Schematic of the AMoRE-I detector system  (Color figure online)}
\end{center}
\label{fig:cryostat}
\end{figure}

The 36 completed modules were stacked in a four-column tower called the AMoRE-I detector tower as shown in Fig.~\ref{fig:modules}. 
We employed thirteen \enCMO{} and five \enLMO{} scintillating crystals, among which six of the \enCMO{} crystals have been used in the AMoRE-pilot experiment. The total mass of the crystals was 6.2 kg, corresponding to 3.0 kg of \Mo{100}, the target isotopes for \zerodbd{} events.
The detector tower had an additional MMC channel without a connection to a crystal. This dummy MMC channel was configured as a sensitive thermometer for the detector tower to maintain stable temperature control~\cite{krwoo}.

The final assembly procedure was performed under an Rn-free atmosphere in a clean-room environment in Y2L. Low-moisture Rn-free air with a dew point lower than $-18\,^\circ$C was supplied throughout the sample preparation periods.

\color{black}

The detector tower was installed at the cryostat used for the pilot phase. Fig.~\ref{fig:cryostat} shows a cross-sectional view of AMoRE-I, including the detector tower, cryostat and shields.
Mass spring dampers that had been used as one of the two-stage vibration mitigation systems~\cite{clee2018} were removed in the AMoRE-I set up.
We added a 5-cm Pb layer to provide a total Pb thickness of 20\,cm for the external shield. 
A muon-veto system, with 5-cm-thick plastic scintillators and photomultiplier tubes, was installed outside the Pb shield to provide nearly 4$\pi$ coverage. 
In addition, a 30-cm-thick polyethylene (PE) layer and a 2.5-cm-thick borated PE layer were installed around the plastic scintillators as neutron shields. We attached an additional 1-cm-thick layer of boric acid powder to the outer vacuum jacket (OVC) of the cryostat.
The AMoRE-I experiment has been running in Y2L at a depth of 700\,m since August, 2020. 

\begin{table*}[b]
\caption{Data obtained for signals from crystals used in the AMoRE-I experiment}
\begin{threeparttable}
\begin{tabular}{llllllll}
\hline
Crystal & Type & Mass & \multicolumn{2}{l}{Pulse height} & \multicolumn{2}{l}{Rise time} \\
No. & & (g) & ($\upPhi_0$) &  & (ms) & \\
 & & & 10 mK & 20 mK & 10 mK & 20 mK \\ \hline
1\tnote{a} & \CMO & 390 & 0.889 & 0.424 & 1.80 & 1.01\\
2\tnote{a} & \CMO & 353& 1.06 & 0.484 & 1.63 & 0.91\\
3\tnote{a} & \CMO & 340& 0.176 & 0.047 & 1.31 & 0.76 \\
4 & \CMO & 358 & 0.764 & 0.383 & 2.46 & 1.25\\
5 & \LMO & 312 & 0.858 & 0.234 & 3.08 & 1.66\\
6 & \LMO & 308 & 0.564 & 0.163 & 2.61 & 1.26\\
7 & \LMO & 312 & 1.082 & 0.228 & 3.67 & 1.80\\
8\tnote{a} & \CMO & 256& 1.26 & 0.601 & 2.02 & 1.02\\
9\tnote{a} & \CMO & 352& 0.735 & 0.352 & 1.74 & 0.97 \\
10 & \LMO & 378 & 0.412 & 0.073 & 2.97 & 1.52\\
11 & \CMO & 426 & 0.669 & 0.313 & 1.70 & 0.98\\
12 & \CMO & 358 & 0.691 & 0.333 & 1.83 & 1.04\\
13 & \CMO & 354 & 0.956 & 0.392 & 2.33 & 1.17\\
14 & \LMO & 300 & 0.702 & 0.177 & 2.74 & 1.39\\
15\tnote{a} & \CMO & 473 & 1.07 & 0.505 & 1.31 & 0.68\\
16 & \CMO & 373 & 0.769 & 0.362 & 2.51 & 1.31\\
17 & \CMO & 356 & 0.736 & 0.351 & 1.84 & 1.00\\
18 & \CMO & 358 & 0.595 & 0.293 & 1.55& 0.90\\ \hline
\end{tabular}
\begin{tablenotes}
\item{a} used in AMoRE-pilot
\end{tablenotes}
\end{threeparttable}
\label{table:crystal}
\end{table*}

\section{Results and Discussions}

The lowest temperature of the cryostat with the dilution refrigerator was 9\,mK during this phase of the experiment. We operated the system during a commissioning period of approximately four months before the new additions to the muon veto system and the neutron shields were complete and ready for the full AMoRE-I. 
Two months of the commissioning period were spent setting up measurement apparatuses for long-term data collection, such as temperature controls, the data acquisition system, the MMC sensors (i.e., injecting persistent currents into the MMC sensors and tuning the SQUIDs), and the reference signals for the stabilization heaters~\cite{dhkwon2020}. During the remainder of the commissioning period, a few calibration runs were carried out at a few different temperatures with a Th source placed in the space between the OVC and the external lead shield. In this report, we present the detector performance and characteristics measured during the commissioning runs. 

All the MMC sensors in AMoRE-I were operated for simultaneous heat and light detection during the calibration runs conducted during the commissioning period. 
Table~\ref{table:crystal} lists the pulse heights and rise-times for 2615-keV $\gamma$ signals from the 18 crystal modules measured at 10\,mK and 20\,mK. 
The signal amplitudes varied with the crystal type and temperatures. 

At 10\,mK, the pulse heights of the phonon signals for all channels except one were in the range of 0.6--1.3\,$\mathrm{\Phi_0}$ corresponding to about 30--60 $\mu$K/MeV sensitivity where $\mathrm{\Phi_0}$ is the magnetic flux quantum, the unit of the SQUID output. The MMC circuits of the heat channel for Crystal No.03 had a ground short and could not support a large current. This channel was found to be inoperable after the commissioning runs were conducted. 
The pulses measured at 20\,mK were smaller in amplitude than those measured at 10\,mK. With increasing temperature, the pulse heights of the \LMO{} crystals were reduced more than those of the \CMO{} crystals. 
The \LMO{} signals in the heat channel generally had longer rise-times than the \CMO{} signals. 
Here, the rise-time is defined as the time interval between 10\% and 90\% of the pulse height in the rising part of the waveform. 

\begin{figure}[t]
\begin{center}
\includegraphics[width=0.75\linewidth, keepaspectratio]{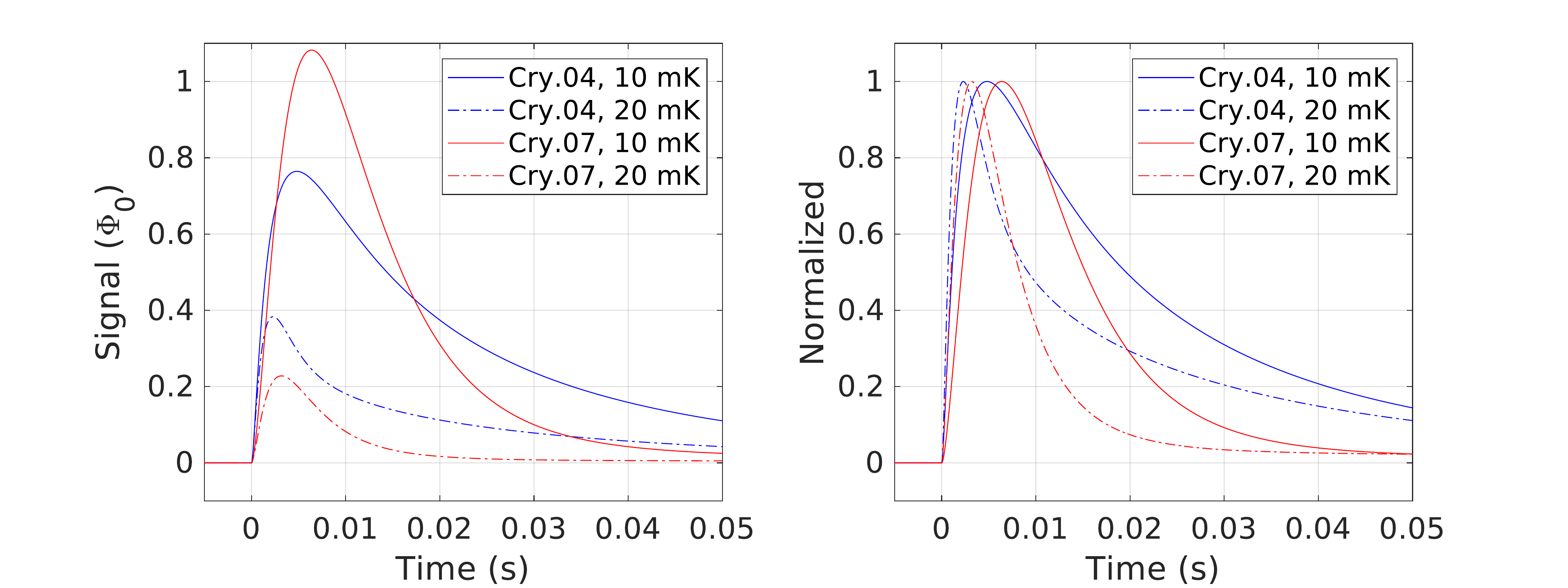}
\caption{ Averaged signals of 2615-keV $\gamma$ events measured in the heat channel at 10 mK and 20 mK for one of each \enCMO{} and \enLMO{} modules in units of one flux quantum ({\it left}) and normalized to the pulse heights ({\it right}). (Color figure online)}
\label{fig:template}
\end{center}
\end{figure}

Fig.~\ref{fig:template} shows the averaged signal waveforms of the 2615-keV $\gamma$ events measured in the heat channel from one \CMO{} crystal and one \LMO{} crystal at 10\,mK and 20\,mK.  The temperature dependence of the pulse amplitudes is noticeably stronger for the signals from \LMO{} than for those from \CMO. The pulse height of the signal from \CMO{} at 20 mK is approximately half that at 10\,mK and approximately 0.2 of that from \LMO{}. We attribute this difference in the temperature dependence of the pulse amplitudes to the athermal heat flow being more dominant in the \CMO{} crystal than in the \LMO{} crystal. We found in our previous studies on using the detector model of heat flow for thermal calorimetric detection~\cite{iwkim2017,gbkim2014} that the heat flows into the phonon collector film on the crystal surface is caused by interactions of thermal and athermal phonons in the absorber crystal with electrons in the metal film. 
Considering a general case of particle detection with a dielectric absorber, initial athermal phonons at 20-50\,K are free from anharmonic downconversion~\cite{tamura1985,maris1993}, and travel around the absorber crystal with a sufficiently long lifetime to hit a metal film on the surface and transfer their energy to the electrons in the film~\cite{yhkim2004}.  
The thermal process is affected by the thermal conductance associated with electron-phonon interactions in the metal film and the heat capacity of the crystal. Both physical quantities have strong temperature dependence. 

Fig.~\ref{fig:template} shows that the \CMO{} signals are affected by the athermal process more strongly than the \LMO{} signals. 
When the thermal process is the dominant heat flow process, the signal pulse height should be inversely proportional to the heat capacity of the absorber. The pulse rise-time should also be a temperature-dependent parameter because this parameter is related to the heat capacity divided by the thermal conductance. Both the pulse heights and rise-times of the \LMO{} signals exhibit a stronger temperature dependence than the \CMO{} signals. This result supports the interpretation that  the thermal process is the dominant heat flow process in \LMO. 

\begin{figure}[t]
\begin{center}
\includegraphics[width=0.6\linewidth, keepaspectratio]{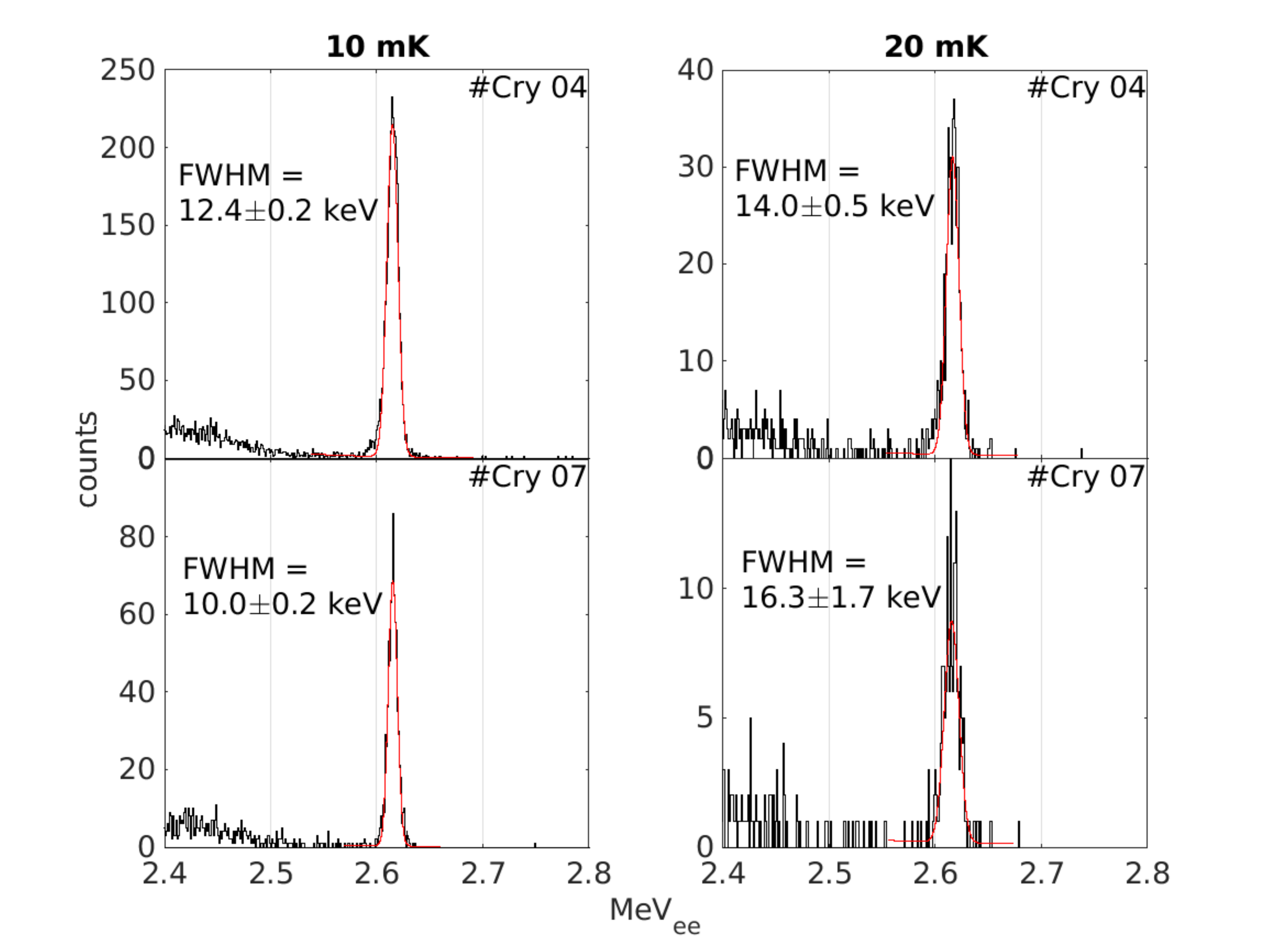}
\caption{
Energy resolution of 2.615-MeV $\gamma$ peaks measured with detector modules of \CMO{} and \LMO{}
at 10 mK and 20 mK. An unbinned likelihood analysis was performed to determine the resolutions.
}
\label{fig:resolution}
\end{center}
\end{figure}

The energy spectra measured by the two detector modules at 10\,mK and 20\,mK are shown in Fig.~\ref{fig:resolution}. 
Generally, the detectors produced higher energy resolutions at 10\,mK than at 20\,mK because of the correspondingly larger pulse amplitudes of the signals.
Considering pulses with similar heights, higher energy resolution was produced by the \LMO{} crystals than by the \CMO{} crystals. In previous studies on \CMO, it was inferred that the event location in the crystal influenced the shape and amplitude of the phonon signals~\cite{gbkim2017,gbkim_ahep2015}. This position dependence, which is strongly associated with the athermal phonon process, became one of the major reasons for the energy resolution limit~\cite{pilot_result}. 
However, energy resolution degradation was apparent at lower temperatures in the \LMO{} detectors as is noticeable in Fig.~\ref{fig:resolution}.

Moreover, the pulse amplitudes of the light signals from all the modules were sufficient to obtain high discrimination power (DP) to identify alpha backgrounds from \bg{} events in the crystals. 

For the two detector modules used to obtain the signals shown in Fig.~\ref{fig:template}, DP values of 7.8 and 5.3 were obtained using \enCMO{} and \enLMO crystals, respectively, where the unitless DP is defined as $(\mu_{\beta} - \mu_{\alpha})/(\sigma_{\alpha}^2 + \sigma_{\beta}^2)^{0.5}$, where $\mu_i$ and $\sigma_i$ are the mean and the standard deviations, respectively, of the relative amplitude ratios between light and heat signals, and the subscript parameters denote $\alpha$ and \bg{} events, respectively. Similar DP values were obtained for the 20\,mK measurements. 
Moreover, the pulse shape (i.e., the rise-time) of the phonon signals corresponded to DP values of 13.3 and 8.7 for the phonon signals of the \enCMO{} at 10\,mK and 20\,mK, respectively. This result indicates that pulse shape discrimination (PSD) is possible for phonon signals from \CMO{} crystals.
However, the phonon signals of all the \LMO\ crystals resulted in DP values in the range of 0.9--3.0. 

\section{Conclusions and plan}

AMoRE-I was started with all the sensors being tuned for a long-term high-resolution detection  for a \zerodbd{}-search experiment. 
The commissioning runs verified that the modules with the two types of molybdate crystals 
provide reasonable detector performance in terms of the energy resolutions and DP values. The calibration runs also suggested that the lower measurement temperature corresponds to higher energy resolution.

As we plan to maintain AMoRE-I in stable operation for long-term data collection, we set the temperature of the detector tower to slightly above 10\,mK. The cryogenic system with the dilution refrigerator has a cooling power of 1.3\,$\mu$W at 12\,mK. We set the cooling power of the control budget to maintain stability over the year-long operation, thereby canceling out any instabilities resulting from fridge operations or utility supplies. 

Moreover, we installed a two-stage temperature-control system using a dummy MMC channel operating in sensitive thermometer mode~\cite{krwoo}. This temperature control can maintain the temperature of the detector tower within an RMS fluctuation of 0.5\,$\mu$K.      
After completing the commissioning periods, we have maintained the detector tower at 12\,mK for science runs for long-term data collection. 

The next generation of the project, AMoRE-II is being prepared to search for \zerodbd{} of \Mo{100} using a 100-kg isotope mass. The development of this large-scale detector is based on the results of AMoRE-I. 

\begin{acknowledgements}

This research is supported by Grant no. IBS-R016-A2, IBS-R016-D1 and NRF-2021R1A2C3010989. 
\end{acknowledgements}

\end{document}